*The study of variability in engineering design – an appreciation and a retrospective*
by
TP Davis

Dept. of Statistics, University of Warwick

*Abstract:*

We explore the concept of parameter design applied to the production of glass beads in the manufacture of metal-encapsulated transistors. The main motivation is to complete the analysis hinted at in the original publication by Jim Morrison in 1957, which was an early example of discussing the idea of transmitted variation in engineering design, and an influential paper in the development of analytic parameter design as a data-centric engineering activity**.** Parameter design is a secondary design activity focussed on selecting the nominals of the design variables, to simultaneously achieve the required functional output, with minimum variance.

*1.     Introduction*

The concept of engineering design robustness has been well established since the work of Japanese engineer Genichi Taguchi became known in the western economies in the 1980s e.g., Taguchi and Wu (1985) and subsequently Taguchi (1986).

Design robustness is formulating a design with a functional response that is as immune as possible to so-called noise factors[1].

The challenge of design robustness has been around for a long time and was uppermost in the minds of 18th century engineers, for example John Harrison in his design of a clock that could keep time despite humidity, and the rolling motion of a ship, to solve the longitude problem, and Josiah Wedgewood in making good quality china pieces, despite variability of temperature inside his kilns, although the word robustness was not common parlance at that time.

Many decades prior to the awareness of Taguchi's work a seminal paper was published by Jim Morrison[2] who addressed a particular type of robustness problem, that of *transmitted variation*. At the time Morrison was working as a glass engineer for the British Thomson-Houston company in Rugby UK, [Morrison (1957)]. This early paper addressed a key problem of robustness associated with the manufacture of encapsulated transistors.

*2.     The glass beads problem*

This problem arose in the early days of semiconductors, whereby development work was proceeding on a new range of metal-encapsulated transistors. The envelope comprised a metal disk to which the germanium or silicon wafer would be attached and a flanged metal "top hat" which would be welded to the disk to complete the enclosure. The electrical connections to the transistor consisted of two fine wires which were sealed into flanged holes on the disk by fusing glass beads to form a hermetic seal. (see Figure 1). This problem provided the motivation for the 1957 paper.

---

[1] Noise factors are sources of variation (either environmental variables or manufacturing variation in design variables) that affect the response. Noise factors are variables that the design engineer cannot control or chooses not to control. Robustness is the task of achieving a functional response on target with minimal variation around the desired target, despite the presence of these noise factors.

[2] I knew both Genichi Taguchi, and Jim Morrison, and learned much from them both.

Although this case study is not recent, it *is* modern. Its main teaching point as we will see, is that the "obvious" (to the uninitiated) solution to this simple problem is wrong, which motivates application of Morrison's method to more complicated problems, whereby the solution will not in any sense be obvious.

It is worth noting that the small glass beads are not just *parts* but are *part of something* (a bigger system – e.g., the complete circuit hardware). Often the performance of an entire system relies heavily on the performance of the smallest parts, for example, in the case of the 1986 Challenger disaster – see Feynman (1986).

After the 1957 paper surfaced (due in large part to the work of George Box and his colleagues at the Center for Quality and Productivity Improvement at the University of Wisconsin[3]), Morrison published a coda to the 1957 paper, [Morrison (1998)], where he introduced the term *variance synthesis* and described his general approach as *statistical engineering*[4]. To quote directly from the 1998 paper "the concept of transmitted error was implicit in the 1957 article, yet it remained neglected and undeveloped until the early 1980s when Taguchi and Wu (1985) introduced off-line quality control and Box and Fung (1986) showed that in suitable circumstances, the error transmission formula provided a better approach…[than those proposed by Taguchi and Wu]" Additionally the 1998 paper gave more detail on how the glass beads were made, which we repeat here verbatim to provide context for the study.

> "Because of the low volume of this special glass, the glass was melted in a small pot and the longer glass tubes from which the beads were cut, were then "drawn" by hand. A skilled glass blower would gather a quantity of molten glass on the end of a blow pipe and would "marver" it to a cylindrical shape on a flat metal plate after putting a puff of air into the gather to keep it hollow, A co-worker would then stick a "punty" on the opposite end of the gather and the two would then walk backwards away from each other along a tube-drawing alley. A third colleague armed with a board would fan the glass to cool it until the tubing was thought to be the right size[5]. The tubing was then cut up into lengths of about 1m prior to shipping to the customer".

There were several types of glass beads mentioned in the 1957 paper: A standard wall bead, a thick wall bead, and a triple-bore bead, but following Morrison (1957), the focus of our discussion is on the standard- wall bead – see Figure 1.

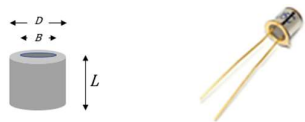

**Figure 1:** a diagram of a single bore bead together with a metal-encapsulated transistor. The glass bead (left figure) ends up inside the "top hat" enclosing the transistor (right figure).

---

[3] Indeed, it was George Box's work at the Center for Quality Improvement at the University of Wisconsin in the mid 1980's that drew this author's attention to the 1957 paper. For a collection of much of the Center's work in this area and at this time see the collected works in Box (2000)
[4] Morrison introduced the term *Statistical Engineering* in some of his earlier publications.
[5] This account explains why variability was a concern in the production of these beads! A YouTube video on making glass cylinders can be found at https://www.youtube.com/watch?v=-eFAeHuGaG4

The volume of a single bore bead is given by the following *objective function*[6].

$$V = \frac{\pi(D^2 - B^2)L}{4} \tag{1}$$

There are three relevant dimensions (the outer diameter ($D$), the inner bore ($B$) and the length($L$)) and these are, collectively the *design variables*. Table 1 shows the measurements of these dimensions taken on a sample of 30 beads, carefully measured with a measurement microscope.

| Bead dimension | sample mean ($\mu_{x_i}$) | sample variance ($\sigma_{x_i}^2$) |
|---|---|---|
| $D$ | $\mu_D = 1.69 mm$ | $\sigma_D^2 = 0.00125\ mm^2$ |
| $B$ | $\mu_B = 0.625 mm$ | $\sigma_B^2 = 0.00254\ mm^2$ |
| $L$ | $\mu_L = 1.92 mm$ | $\sigma_L^2 = 0.00536\ mm^2$ |
| Bead volume ($V$) | $\mu_V = 3.72\ mm^3$ | $\sigma_V^2 = 0.0617\ mm^6$ |

**Table 1:** The mean dimensions and variances of a sample of thirty beads, as reported in Morrison (1957). The volumes in Table 1 were estimated from the volume equation (1) and were calibrated by estimating the volumes by weighing (ignoring any variation in glass density). The volume estimates derived from the glass density, were on average lower than those calculated geometrically, primarily due to the beads not being perfectly circular, and some of the beads losing some material by being chipped. (a failure mode which we ignore in this paper).

There were two distinct failure modes which hampered high quality production of the beads. Firstly, the glass beads would crack under temperature gradients due to the differences in thermal properties of the glass and the metal of the mounting surface that was being sealed. A so-called *one-sided* failure mode (Clausing & Frey 2005). Secondly, at the pre-production stage, a high failure rate was being experienced in the glass/metal sealing process because the volume of the glass in production was too variable. For example, ($a$) If the volume of the glass bead was too great, the fused glass would spread across the disk and impede the mounting of the wafer at the next production stage, and ($b$) if the volume was too small, surface tension would pull the glass to one side leaving a gap in the seal meaning the seal wasn't airtight. This is a *two-sided* failure mode illustrated in Figure 2, and is the main focus of this paper.

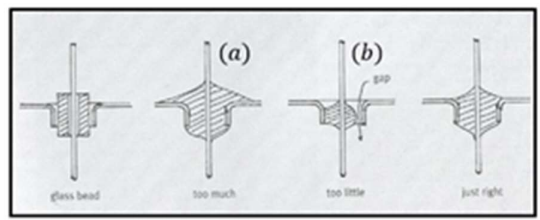

**Figure 2:** an illustration of the two-sided failure mode due variation in glass volume, as sketched by Jim Morrison.

---

[6] Sometimes called a *transfer function*.

Both failure modes are caused by noise factors – in the glass cracking case, the noise factor is an environmental variable (temperature), and the variability in volume the noise factor is variation in the design variables which is *transmitted* to the volume.

The countermeasure for the glass cracking was to make the glass from a special borosilicate composition whose expansion characteristics matched those of the metal it was sealing, so that the seal would be stress free (in other words *robust* to temperature).

The variance that is transmitted from an input variable $x$ say to an output variable $y$ say is given by the expression $\sigma_y^2 \approx \frac{\partial y}{\partial x} \sigma_x^2$, where the derivative of the objective function is evaluated at the point of interest in the $x$ space (usual the nominal or target value, such that when $x$ takes this value the required output value $y$ is achieved. The expression is of course exact if the relationship between $y$ and $x$ is linear, and will provide an adequate approximation if $y = f(x)$ is roughly linear in a near-neighbourhood of $x$.[7]

When there is more than one design variable, there usually is, say $\{x_i\}_{i=1}^k$. In the glass bead example, there are three design variables, $\{x_i\}_{i=1}^3 = \{D, B, L\}$, and so the variance transmission formula is (assuming the $x_i$s are independent[8])

$$\sigma_y^2 = \sum_{i=1}^{3} \left(\frac{\partial y}{\partial x_i}\right)^2 \cdot \sigma_{x_i}^2 \tag{2}$$

where the derivatives are calculated at the nominals for the $x_i$s. we take the nominal values as the mean values in Table 1. The partial derivatives are easily derived and so equation (2) (with $V$ taking the role of $y$) becomes

$$\sigma_V^2 = \left(\frac{\partial V}{\partial D}\right)^2 \sigma_D^2 + \left(\frac{\partial V}{\partial B}\right)^2 \sigma_B^2 + \left(\frac{\partial V}{\partial L}\right)^2 \sigma_L^2 \tag{3}$$

$$\sigma_V^2 = \left(\frac{\pi D L}{2}\right)^2 \sigma_D^2 + \left(\frac{\pi B L}{2}\right)^2 \sigma_B^2 + \left(\frac{\pi (D^2 - B^2)}{4}\right)^2 \sigma_L^2 \tag{3a}$$

so $\quad \sigma_V^2 = (25.98 \times 0.00125) + (3.55 \times 0.00254) + (3.75 \times 0.00536) \tag{3b}$

and finally,

$$\sigma_V^2 = 0.0325 + 0.0090 + 0.0201 = 0.0617 \text{mm}^6 \tag{3c}$$

Note that each derivative has units mm[4] So that these equations are dimensionally consistent. Observe that it is the diameter $D$ that accounts for most of the variability transmitted to $V$, and not $L$ even though $\sigma_L^2 > \sigma_D^2$. This is because the gradient in the direction of $D$ (25.98) is much greater than in the direction $L$ (3.75). This example shows that even in such a simple case the "obvious" solution of attacking the design variable with the largest variance is wrong. It can only be speculated as to how intuition might let us down in more complicated examples.

Before this analysis was done by Jim Morrison, The British Thomson-Houston Company Ltd. (which was part of GE) was planning to spend a lot of money investing in a new cutting machine to improve the accuracy of the cut length ($L$) of the beads.

---

[7] Morrison (1957) suggests that equation (2) will be adequate as long as the standard deviation is less than 20% of the mean.
[8] In the 1957 account Morrison indicates that there is "a slight degree of correlation" between $B$ an $D$. Unfortunately, the original data is no longer available to check the value of this correlation, but we investigate the consequences of such a correlation, based on plausible assumptions, further on in the paper.

It turned out that the reduction in the variance of the diameter was easier to achieve than reduction in the variance of the length because it was observed that $D$ only varied gradually along a cut length of glass tubing of ~1m in length.

By cutting these longer 1m lengths into shorter pieces of ~10cm and producing batches of beads from these shorter lengths, the variability of $D$ within a batch of these 10cm lengths was reduced which in turn resulted in a large reduction in the variance of the glass bead volume. This clever control plan based on sorting the shorter batches bypassed the need for new cutting equipment to control cut length $L$, since the glass beads were easily batched according to their origin from the shorter pieces[9]. The glass bead example nicely illustrates the importance of analysing transmitted variation in making the right engineering and manufacturing decisions to improve quality and avoid unnecessary cost.

The analysis of transmitted variation was called *tolerance design* by Taguchi (1986). Jim Morrison came to call this *variance synthesis.* [Morrison (1998)].

Tolerance design is a *tertiary* design activity – The *primary* design activity is establishing the basic design concept to deliver the functionality as required by the customer or end-user.

A *secondary* design activity is choosing the correct nominals for the design variables so that the objective target is achieved. If the secondary design activity *also* has the objective of improving design robustness (minimizing transmitted variation), then Taguchi called this *parameter design*.

In his 1957 paper Jim Morrison didn't pursue a parameter design solution for the glass beads, although a careful reading of the paper did show that he had thought about it: "In situations in which bead volume is a critical factor, the designer can use this analysis as a guide in determining the most suitable proportions of bead for a given application".

Bisgaard and Ankemann (2005) claim that this is the first reference to parameter design in the literature. So, we look at parameter design in some detail now.

### 3. A parameter design study of the glass beads

The parameter design problem for the glass beads can be stated as minimize the transmitted variation given by (3) subject to $V = 3.72mm$ (equation (1)). Here we use the Solver function in MS Excel® to determine the solution.

We make some realistic assumptions before proceeding; firstly, we assume a constant coefficient of variation $\frac{\sigma_{x_i}}{\mu_{x_i}}$ for each of the 3 design variables $\{x_i\}_{i=1}^{3} = \{D, B, L\}$ since in engineering and manufacturing applications it is often true that the variability and the mean are *linked*. A general link function is $\sigma_{x_i} \propto \mu_{x_i}^p$, where a constant coefficient of variation corresponds to the case $p = 1$. But other values of $p$ may be appropriate (see Box (1988)). We begin by assuming that $p = 1$. The coefficients of variation can be calculated directly from Table 1. Secondly, we add the constraint that the bore dimension $B$, must be in the interval (0.6$mm$–0.65$mm$), so that the thin wires can pass through the bead easily. Also, the Diameter $D$ clearly needs to be greater than the Bore, so $D>B$, and it cannot be too large otherwise the bead will not locate into the "top hat", so we arbitrarily set the constraint $D$≤2$mm$.

We take as the starting values or initial design specification the measurements given in Morrison (1957). These are summarized in Table 2.

---
[9] It may be tempting to think that these days sorting to achieve high quality is not required. But even in high precision manufacturing sorting can be effective, e.g., in the fitting of fan blades to modern jet engines.

| Dimension | Initial design specification ($\mu_{x_i}$) | Variance ($\sigma_{x_i}^2$) | Assumed coefficient of variation $\frac{\sigma_{x_i}}{\mu_{x_i}}$ | Variance transmitted to V from eq (3) |
|---|---|---|---|---|
| D | 1.69mm | 0.00125mm² | 0.02092 | 0.0325mm⁶ |
| B | 0.625mm | 0.00254mm² | 0.00806 | 0.0090mm⁶ |
| L | 1.92mm | 0.00536mm² | 0.03813 | 0.0201 mm⁶ |
| Volume (V) | 3.72mm³ | | | 0.0616mm⁶ |

**Table 2:** The initial design specification prior to conducting parameter design, with the resulting transmitted variation.

After running the Excel solver with the constraints as specified, the results are as given in Table 3. Note the coefficients of variation are the same in each case per our assumption.

| Dimension | Design specification after parameter design ($\mu_{x_i}$) | variance ($\sigma_{x_i}^2$) | coefficient of variation $\frac{\sigma_{x_i}}{\mu_{x_i}}$ | Variance transmitted to V |
|---|---|---|---|---|
| D | 2.0mm | 0.00175mm² | 0.02092 | 0.0293mm⁶ |
| B | 0.6mm | 0.00234mm² | 0.00806 | 0.0035mm⁶ |
| L | 1.30mm | 0.00246mm² | 0.03813 | 0.0201 mm⁶ |
| Volume (V) | 3.72mm³ | | | 0.0529mm⁶ |

**Table 3:** The parameter design solution. Note that the coefficients of variation are as in Table 2 per our assumption i.e., $\sigma_{x_i} \propto \mu_{x_i}$

Note that comparing the post-parameter design in Table 3 with the initial specification in Table 2, the transmitted variation has reduced from $0.0616$ to $0.0529$ or about $14\%$. This gain is equivalent to reducing the variance of $D$ in the initial specification by about $20\%$. Note in Table 3 the changes in the nominal values for $D, B \& L$ compared to Table 2. Figure 3 illustrates the parameter design solution graphically.

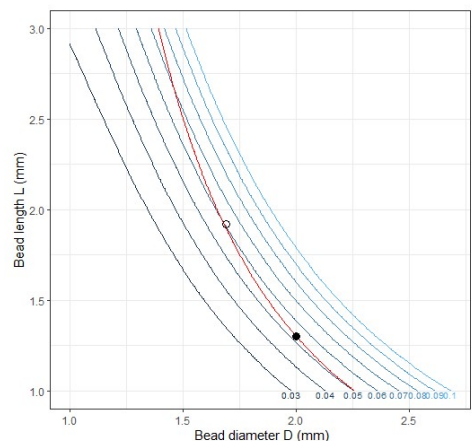

**Figure 3:** contours of transmitted variation from bead dimensions $D$ and $L$ to volume $V$ under the assumption that $\sigma_{x_i} \propto \mu_{x_i}$. The red line corresponds to $V = 3.72$mm. In this plot the Bore diameter $B$ is set to its optimal value of 0.6mm. The open dot is the initial specification (slightly off the red line because initially $B \neq 0.6$), and the solid dot is the parameter design solution.

Note from Figure 3 that the optimal value of the outer diameter D is on the edge of its constraint, so if this constraint could be relaxed e.g., by using a larger "top hat", a better solution (i.e., a further reduction in $\sigma_V^2$) could be realised[10].

It has long been recognised, that the importance of the assumption regarding the relationship between $\mu_{x_i}$ and $\sigma_{x_i}$ is crucial in parameter design. For example, if we repeat the previous analysis, but now assume that the variances for the bead dimensions are fixed at the values in Table 1, (i.e., $p = 0$) the parameter design solution is as in Table 4.

| Dimension $x_i$ | Parameter design nominal($\mu_{x_i}$) | Variance assumed fixed ($\sigma_{x_i}^2$) | Variance transmitted to $V$ |
|---|---|---|---|
| $D$ | 1.74mm | 0.00175mm$^2$ | 0.0294mm$^6$ |
| $B$ | 0.6mm | 0.00234mm$^2$ | 0.0071mm$^6$ |
| $L$ | 1.77mm | 0.00246mm$^2$ | 0.0236 mm$^6$ |
| Volume ($V$) | 3.72mm$^3$ |  | 0.0601mm$^6$ |

Table 4: parameter design solution assuming that the variances of the bead dimensions are fixed at the values given in Table 2.

Note from Table 4, that the proposed nominals for $D$ and $L$ are quite different to the previous solution. This sensitivity of the parameter design method was acknowledged by Morrison (1957) and was discussed more extensively by Box & Fung (1994).

The contour plot under the assumption of fixed variances for $D, B$ & $L$ is shown in Figure 4.

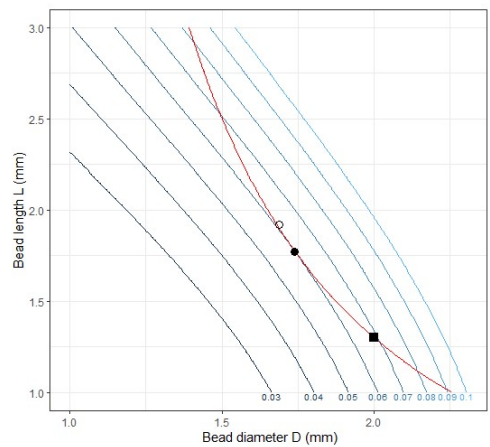

Figure 4: Contours of the transmitted variation now assuming *fixed* variances for the bead dimensions, showing the new parameter design solution (solid dot), compared to the previous solution under the assumption of constant coefficient of variation (the solid square, now with $\sigma_V^2 = 0.068470$). The initial design specification is shown by the open dot.

Note that the contours in Figure 4 are oriented differently compared to those in Figure 3, illustrating the sensitivity of parameter design to underlying assumptions, to the extent that optimal value of the outer diameter $D$ is now well within the imposed constraint. Clearly in this case, and in general, the functional relationship between $\sigma_{x_i}$ and $\mu_{x_i}$ needs to be carefully established.

---

[10] The diameter of a glass tube is a function of the speed at which the glass is drawn.

Additionally, to the functional relationship between $\sigma_{x_i}$ and $\mu_{x_i}$, we should also check for the effect of a lack of independence between the design variables. Morrison (1957) hints that variables $B$ and $D$ were slightly correlated (larger bores tended to result in larger outer diameters). If we denote the correlation between $B$ and $D$ by $\rho_{DB}$ an additional term must be added to the variance transmission formula, (3), namely

$$2\frac{\partial V}{\partial B} \cdot \frac{\partial V}{\partial D} \sigma_B \sigma_D \rho_{DB} = \frac{\pi^2 BDL^2}{2} \cdot \sigma_B \sigma_D \rho_{DB}.$$

Note that with this extra term equation (3) is still dimensionally consistent.

The results in Table 5 illustrate the effects of a weak correlation between $B$ and $D$ (say $\rho_{BD} \leq 0.3$). Note that the optimal value for the design variables only change when $p = 0$, and the transmitted variation to $V$ increases with $\rho_{BD}$. The orientation of the contours of transmitted variation do not change with $\rho_{DB}$ so we do not show them here.

| $\rho_{BD}$ | $p = 0$ (i.e., $\sigma_{x_i}$ is fixed) | | $p = 1$ (i.e., $\sigma_{x_i} \propto \mu_{x_i}$) | |
|---|---|---|---|---|
| | $\sigma_V^2$ with $\{D, B, L\}$ | Contribution of $\rho_{DB}$ to $\sigma_V^2$ | $\sigma_V^2$ with $\{D, B, L\}$ | Contribution of $\rho_{DB}$ to $\sigma_V^2$ |
| 0.0 | 0.060079 {1.74,0.6,1.77} | 0.0 | 0.052896 {2,0.6,1.3} | 0.0 |
| 0.1 | 0.062911 {1.76,0.6,1.73} | 0.0018 | 0.054926 {2,0.6,1.3} | 0.0020 |
| 0.2 | 0.065633 {1.78,0.6,1.69} | 0.0036 | 0.056956 {2,0.6,1.3} | 0.0041 |
| 0.3 | 0.068259 {1.80,0.6,1.65} | 0.0054 | 0.058986 {2,0.6,1.3} | 0.0061 |

**Table 5:** the consequences of a correlation between design variables (in this case D and B) Note that when $p = 0$, the optimal values for D and L depend on $\rho_{DB}$, due to the orientation of the contours so this highlights that the independence of the design variables needs to be considered in parameter design studies in addition to the considerations emphasised in Box & Fung (1994).

One other approach to deal with correlated variables in a parameter design study is to express the objective function in dimensionless variables using an application of Buckingham's "Pi theorem" (e.g., see Shen *et al* 2014). For the objective function (1), if we take as our dimensionless variables, $\pi_0 = \frac{V}{B^3}$, $\pi_1 = \frac{D}{B}$, and $\pi_2 = \frac{L}{B}$, then $\pi_0 \propto (\pi_1^2 - 1) \cdot \pi_2$. The dimension of the problem is reduced by one. Note that the two variables which exhibit a correlation now appear as a ratio. Grove and Davis (1992) show how dimensionless variables help with parameter design by exploring this idea while analysing Taguchi's celebrated and widely taught Wheatstone Bridge problem (see Box, 2000, Chapter E.3), albeit in that case to deal with an interaction between design variables rather than a correlation.

Determining a value for $p$ from the summary data for $\{D, B, L\}$ regarding the three types of beads discussed in the 1957 paper is inconclusive, but there is some evidence that for $D$ and $B$, $p = 0$, while for $L$, $p = 1$. To conclude the parameter design discussion, we investigate the effect that this hybrid assumption has on the parameter design solution.

Firstly assuming $\rho_{DB} = 0$ to allow for direct comparison to the solutions illustrated in Figures 3 and 4, the resulting contour plot is shown in Figure 5. The optimal value for $\sigma_V^2$ is now $0.44830$.

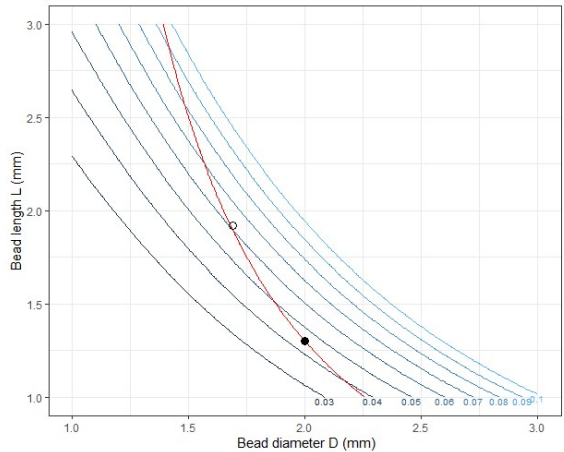

**Figure 5:** The parameter design solution with $\rho_{DB} = 0$ and $p = 0$ for $D$ & $B$, and $p = 1$ for $L$. The optimal values for $\{D, B, L\}$ are $\{2.0, 0.6, 1.3\}$, with $\sigma_V^2 = 0.044830$. As in the previous figures the open dot is the initial bead specification and the parameter design solution by the solid dot.

Note that in Figure 5, the orientation of the contours is similar to Figure 3 when it was assumed that $p = 1 \,\forall \{D, B, L\}$, so it is the variance assumption on $L$ which dictates the re-orientation of the contours observed in Figure 4.

Exploring this hybrid assumption now for the range of values for $\rho_{DB}$ that were considered previously we find the results in Table 6. Note that although the optimal values of $\{D, B, L\}$ do not alter from the assumption of $p = 1$ for each design variable, the optimal value for $\sigma_V^2$ is improved from the previous results.

| $\rho_{BD}$ | Optimal $\{D, B, L\}$ for $p = \{0,0,1\}$ | $\sigma_V^2$ | Contribution of $\rho_{DB}$ to $\sigma_V^2$ |
|---|---|---|---|
| 0.0 | $\{2.0, 0.6, 1.3\}$ | 0.044830 | 0.0 |
| 0.1 | $\{2.0, 0.6, 1.3\}$ | 0.046615 | 0.0018 |
| 0.2 | $\{2.0, 0.6, 1.3\}$ | 0.048403 | 0.0036 |
| 0.3 | $\{2.0, 0.6, 1.3\}$ | 0.050190 | 0.0054 |

**Table 6:** Parameter design solution under the assumption $p = \{0,0,1\}$ for $\{D, B, L\}$ for various values of $\rho_{DB}$.

In summary we see that for the glass bead study it is the assumption regarding the link between $\sigma_L$ and $\mu_L$ that is crucial for this analysis on three counts – the orientation of the contours and the optimal settings for $\{D, B, L\}$ and magnitude of the transmitted variance.

### 4. In retrospect

Morrison's 1957 paper was it seems the first paper to formulate the idea of parameter design although he did not call it that and he did not explore that idea in his paper. Together with the fact that he published the paper in a journal that engineers don't usually read, may explain why Morrison's early ideas weren't picked up in the engineering mainstream until the work of Taguchi became better known. Subsequently George Box and his co-workers at the Center for Quality and productivity improvement at the University of Wisconsin-Madison regularly referenced the 1957 paper in their many publications.

The method was used in the Ford Motor Company (Parry-Jones 1999, Davis 2006) as part of a coordinated strategy to embed robust design into engineering practice and was taught in company training programs.

It is hoped that as more data-centric engineering methods (Girolami 2020), are developed as part of the statistical engineering framework, Morrison's work with suitable modern adaptions will find continue to find its place in current engineering design challenges. Although the glass bead example is simple to understand, the solution to reducing the transmitted variation is not obvious, which has implications for applications with many more design variables. Recent developments in applying parameter design for situations that rely on a computer model (a "digital-twin") to evaluate design alternatives, because an explicit objective function is not available due to complexity and high dimension, is given in Shen (2017).

In his 1998 coda to the 1957 paper Morrison laid out an 11-step process for reducing variability in engineering design. The final step was (my italics)
"The possibility of experimenting with the nominal values of the design parameters ... can be explored *especially if the variances are not constant but alter with changes in the nominal values*."

The analysis of the glass beads problem presented here illustrates step 11 for the glass beads problem and emphasizes the importance of checking the assumptions regarding the way in which the standard deviation and mean of the design variables may be linked.

The purpose of this paper is to recognise and honour the importance of Jim Morrison's paper as a foundational paper for statistical engineering, and to investigate the parameter design solution to the glass bead problem that Morrison envisaged but did not actually do. My one regret is that I did not do this sooner, so that I could have shared this work with him[11].

## 5. Acknowledgments

I would like to thank Dr. Shirley Coleman, of Newcastle University, Jon Bridges of the University of Bradford, and an anonymous referee for making constructive comments on an earlier draft.

## 6. Author contributions

TPD is the sole contributor to this work.

---

[11] Jim Morrison's obituary can be found in *Journal of the Royal Statistical Society Series A, Volume 180, Issue 1 pp 348-350* and online at https://rss.onlinelibrary.wiley.com/doi/full/10.1111/rssa.12271